\documentclass[11pt]{amsart}
\usepackage{amsmath,amsthm,amssymb,amscd, graphicx, wrapfig}
\usepackage[all]{xy}
\usepackage[margin=1.1in]{geometry}
\usepackage[usenames, dvipsnames]{color}
\usepackage[bookmarks=false,colorlinks,linkcolor=blue,citecolor=blue]{hyperref}

\usepackage[T1]{fontenc}

\theoremstyle{plain}

\theoremstyle{definition}

\theoremstyle{remark}
\newtheorem*{rem}{Remark}

\newtheorem*{ack}{Acknowledgments}

\numberwithin{equation}{section}

\newcommand{\R} {{\mathbb R}}                              
\newcommand{\Z} {{\mathbb Z}}                              

\newcommand{\hide}[1]{}

\begin{document}

\title{Unfolding Overlaps of the Exceptional Regular Polytopes}

\author{Satyan L.\ Devadoss}
\address{S.\ Devadoss: University of San Diego, San Diego CA 92110}
\email{devadoss@sandiego.edu}

\author{Matthew Harvey}
\address{M.\ Harvey: The University of Virginia's College at Wise, Wise VA 24293}
\email{msh3e@uvawise.edu}

\author{David Richter}
\address{D.\ Richter: Western Michigan University,  Kalamazoo MI 49008}
\email{david.richter@wmich.edu}

\begin{abstract}
We find explicit ridge unfoldings of the three exceptional 4D polytopes (24-cell, 120-cell, 600-cell) that result in overlaps of their facets. These  failures bring an end to the full classification of regular polytopes with the all-net property.
\end{abstract}


\maketitle
\baselineskip=18pt
\renewcommand\arraystretch{1.3}

%
%
\section{Introduction} \label{s:intro}
\subsection{Historical Framing}

Unfolding polyhedra was popularized by the German Renaissance master Albrecht D\"urer.  His influential book \emph{The Painter's Manual} contains the first recorded example of a \emph{net}: a polyhedron cut along some of its edges and unfolded to a single, non-overlapping, simple polygon in the plane.
In the 1970s, Shephard  \cite{sh1} conjectured that every convex polyhedron admits a net along some edge cuts, a claim that remains tantalizingly open to this day.

For a higher-dimensional \emph{polytope}, its codimension-one faces are \emph{facets} and its codimension-two faces are \emph{ridges}. The analog of an edge unfolding of polyhedron is the \emph{ridge unfolding} of an $n$-dimensional polytope: cutting the polytope along a collection of its ridges where the resulting (connected) arrangement of its facets maps isometrically into an $\R^{n-1}$ hyperplane.  If this map is an embedding (where the arrangement does not overlap in the hyperplane), then the ridge unfolding of the polytope yields a net.
The works of Miller and Pak \cite{mp}, and especially the collected works of Alexandrov \cite{alex1}, serve as wonderful historical readings for higher-dimensional unfoldings.

Our intention is to consider a stronger property than just the discovery of a single net: We say a polytope 
$P$ is \emph{all-net} if every ridge unfolding of $P$ yields a valid net. It would seem imprudent to 
explore all-nets for polytopes when finding even one net for 3D polyhedra remains open. And so, we turn 
our attention to a special class, the \emph{regular polytopes}, the generalizations of the regular 
polygons. In particular, the symmetry group of a regular polytope acts transitively on its flags \cite{hsm}.

\subsection{Unfolding Regular Polytopes}

There are five regular polyhedra in three dimensions, notably the Platonic solids, the culminating proposition in Euclid's Elements. In 2011, using computer calculations, Horiyama and Shoji \cite{hosh} showed that the five Platonic solids are all-net. This is possible to check by hand for the tetrahedron, cube, and octahedron, for which there are few unfoldings, but it is significantly nontrivial for the dodecahedron and icosahedron, for which there are 43,380 distinct unfoldings each. 

For higher dimensions, the regular polytopes were fully classified by Ludwig Schl\"afli in the 1800s, and generalized in a group-theoretic setting by H.S.M.\ Coxeter \cite{hsm} in the 1970s. For dimensions five and greater, there are exactly three classes of regular polytopes: $n$-simplex, $n$-cube, $n$-orthoplex. All such simplicies and  cubes were shown to be all-net, and surprisingly, all such orthoplexes were shown to fail \cite{ddrw, dh}.

Finally, in four dimensions, there are six  regular polytopes: the three from above (simplex, cube, orthoplex) along with three exceptional ones (24-cell, 120-cell, 600-cell). In a lovely paper from 1998, Buekenhout and Parker \cite{bupa} enumerate the number of distinct ridge unfoldings of these regular  4-polytopes (up to symmetry), shown in Table~\ref{t:4D}. We can immediately observe that the first three objects are quite different than the last three, and the unfolding mathematics match: The 4-simplex, 4-cube, and 4-orthoplex  have the all-net property \cite{dh}, and we encourage the reader to explore the interactive software \cite{zhang} by Sam Zhang to create every net of these polytopes. 

\begin{table}[h]
\hrule
\begin{align*}
{\text{simplex}} & \ : \ \ 3 \\
{\text{cube}} & \ : \ \ 261 \\
{\text{orthoplex}} & \ : \ \ 110,912 \\
{\text{24-cell}} & \ : \ \ 6 \ (2^{19} \cdot 5688888889 \ + \ 347) \\
{\text{120-cell}} & \ : \ \ 2^7 \cdot 5^2 \cdot 7^3 \ (2^{114} \cdot 3^{78} \cdot 5^{20} \cdot 7^{33} \ + \ 2^{47} \cdot 3^{18} \cdot 5^2 \cdot 7^{12} \cdot 53^{5} \cdot 2311^3 \ + \ 239^2 \cdot 3931^2) \\
{\text{600-cell}} & \ : \ \ 2^{188} \cdot 3^{102} \cdot 5^{20} \cdot 7^{36} \cdot 11^{48} \cdot 23^{48} \cdot 29^{30}
\end{align*}
\hrule
\caption{The number of distinct unfoldings of the six regular 4-polytopes.}
\label{t:4D}
\end{table}

In this paper, we demonstrate that the three exceptional regular polytopes fail the all-net condition. 
We accomplish this by exhibiting a chain of successively adjacent facets (or, by duality, a path in the edge skeleton
of the dual) in each of the three cases. 
Section~\ref{s:24} focuses on a chain of 14 octahedra in the 24-cell, 
Section~\ref{s:120} shows a chain of 9 dodecahedra in the 120-cell tiled by dodecahedra,
and Section~\ref{s:600} shows a chain of 8 tetrahedra in the 600-cell.
We note that the chain of length 8 for the 600-cell was already
demonstrated in \cite[Theorem 11]{dh}, but we include it here for completeness.
And thus, the full results for all-net unfoldings of regular polytopes can be given in Table~\ref{t:allnet}, where $n \geq 5$.

\begin{table}[h]
\begin{tabular}{cllllll}
3D & \textcolor{black}{Y: tetrahedron}  & \textcolor{black}{Y: cube} & \textcolor{black}{Y: octahedron}  & \textcolor{black}{Y: dodecahedron}  & \textcolor{black}{Y: icosahedron} &  \\
4D & \textcolor{black}{Y: 4-simplex} & \textcolor{black}{Y: 4-cube}  & \textcolor{black}{Y: 4-orthoplex} & \textcolor{red}{N: 120-cell} & \textcolor{red}{N: 600-cell} & \textcolor{red}{N: 24-cell} \\ 
$n$D & \textcolor{black}{Y: $n$-simplex} & \textcolor{black}{Y: $n$-cube} & \textcolor{red}{N: $n$-orthoplex}  & & \\ \\
\end{tabular}
\caption{All-net results for regular polytopes, where $n \geq 5$.}
\label{t:allnet}
\vspace{-.2in}
\end{table}

\subsection{Strategy}

For each case, we exhibit a tuple 
$f_1, \dots, f_n$ of facets, where the first and last facet in this chain intersect when unfolded into three dimensions.
For each $i\in\{1, \dots, n\}$, we have an unfolding function $M_i:\R^4\rightarrow\R^3$ that maps the facet
$f_i$ to a regular 3-dimensional polyhedron (octahedron for the 24-cell, dodecahedron for the 120-cell,
 tetrahedron for the 600-cell).
One verifies all of the following in order to guarantee that our chain results in an unfolding overlap:
\begin{itemize}
\item The facets $\{f_1, \dots, f_{n}\}$ are distinct facets of the polytope.

\item The images $\{M_1(f_1), \dots, M_n(f_{n})\}$ are distinct facets of the polytope.

\item For each $i\in\{1, \dots, n-1\}$, the cells $f_i$ and $f_{i+1}$ share a 2D face (and are therefore adjacent) and the images $M_i(f_i)$ and $M_{i+1}(f_{i+1})$ share a 2D face.

\item At least one point of $M_{n}(f_{n})$ is an interior point of $M_1(f_1)$.
\end{itemize}
All of these are routine calculations in matrix algebra.  

\begin{rem}
The maps $M_2, \dots, M_n$
are uniquely determined by the tuple $\{f_1, \dots, f_n\}$ and the initial choice of $M_1$. 
Each map $M_{i+1}$ can be determined from $M_i$ by the images $M_i(v)$ of several points of
$v$ of $f_i$, as each can be obtained by reflecting vertices of $f_i$ across the plane 
stipulated to be common to both $M_i(f_i)$ and $M_{i+1}(f_{i+1})$, by virtue of regularity.
\end{rem}

\begin{ack}
 We thank Joe O'Rourke for his encouragement, and Robert Webb and his \emph{Stella4D} software for inspiring the unfolding in Figure~\ref{f:120-paper}.  The overlap results in this paper were finished in February 2025. The first author takes full responsibility for the tardiness in bringing this work to our community.
\end{ack}

%
%
\section{The 24-cell} \label{s:24}
\subsection{Overview}

We recall some properties of the 24-cell: it is a 4D convex polytope having 24 identical octahedral facets (also called \emph{cells}),  96 equilateral triangular faces, 96 edges, and 24 vertices. 
Although the 24-cell is self-dual\footnote{Self-duality in the sense that there is an isomorphism of the face lattice which interchanges vertices with facets and edges with triangles.}, in contrast to the regular simplex, there is no canonical self-duality.
The vertices of our 24-cell consist of the 24 points formed when
we apply all possible sign changes and permutations of the four coordinates of the point 
$(2,2,0,0)\in\mathbb{R}^4$.  With these coordinates,
vertices $v$ and $w$ are joined by an edge of the 24-cell if and only if $v\cdot w=4$.
The centroids in $\R^4$ of the octahedral cells are the 24 points with
	\begin{itemize}
	\item 8 points from permutations of $(\pm 2,0,0,0)$
	\item 16 points of the form $(\pm 1,\pm 1,\pm 1, \pm 1)$.
	\end{itemize}
If $C$ is a centroid of an octahedron, then a point $v$ in our vertex set is vertex of this
octahedron if and only if $C\cdot v=4$.  Thus, for example, the vertices of the octahedron
with centroid $C=(2,0,0,0)$ consists of the six points
$$(2,2,0,0),(2,-2,0,0),(2,0,2,0),(2,0,-2,0),(2,0,0,2),(2,0,0,-2).$$
Similarly, the vertices of the octahedron with centroid $C=(1,1,1,1)$ are the points
$$(2,2,0,0),(0,0,2,2),(2,0,2,0),(0,2,0,2),(2,0,0,2),(0,2,2,0).$$
We also note that two centroids $C_i$ and $C_j$ lie in adjacent octahedral cells if and only if $C_i \cdot C_j=2$.  This is evident in the example above.

\subsection{The Unfolded Chain}

We present a chain $\{f_1,f_2,...,f_{14}\}$ of successively adjacent facets
of the 24-cell such that any unfolding of the 24-cell which contains this chain 
is necessarily non-net.  For each $i$, we represent a linear map  $M_i:\mathbb{R}^4\rightarrow\mathbb{R}^3$ by a $4\times 3$ matrix and regard $M_i$ as a component of the 24-cell unfolding along our chain. 

\begin{figure}[h]
\includegraphics[width=.9\textwidth]{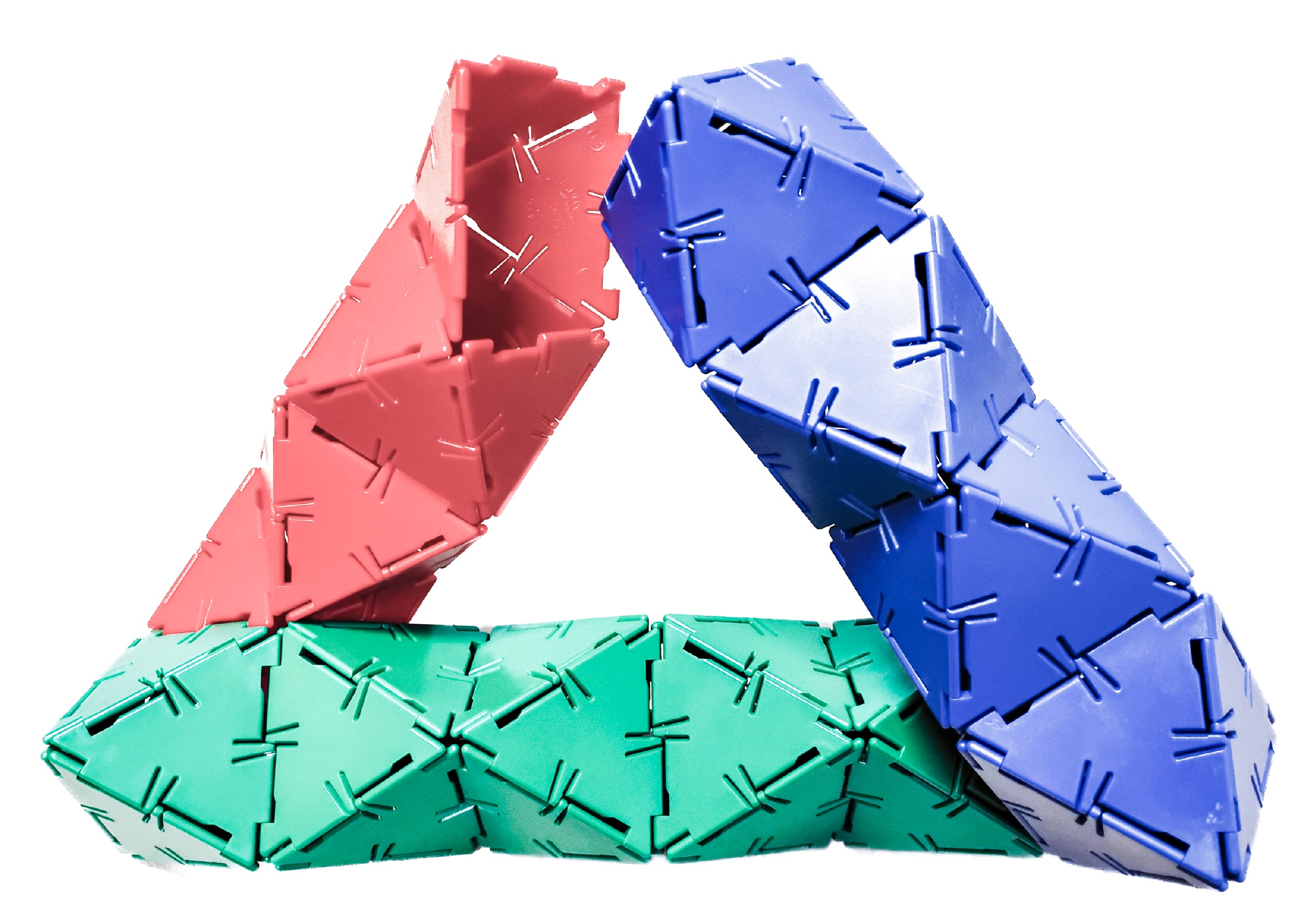}
\caption{Chain of 14 octahedra, unfolded from three toroidal rings.}
\label{f:24-model}
\end{figure}

The octahedra in the 24-cell may be partitioned into four rings of six octahedra, where each ring  folds to a solid torus in 4D space, and each pair of rings is linked as in the Hopf fibration of the 3-sphere.  
Figure~\ref{f:24-model} shows a physical model of a length-14 octahedral chain made from \emph{Polydron} panels. The (green) chain with six cells in the middle folds to a solid torus, and the two appended (blue and red) chains with four cells each would also fold into solid tori if two additional cells are included. However, since these length-four chains already create an overlap, we don't need to complete the two towers on the ends.

\begin{rem}
In our computations, we chose the fifth octahedron in this chain as the center of our unfolding: the matrix product $M_5C_5$ is the origin in $\mathbb{R}^3$ with an augmented coordinate 1 to realize $\mathbb{R}^3$ as an affine subspace of $\mathbb{R}^4$. The remaining matrices $M_i$ were obtained from this one.
\end{rem}

\subsection{The Overlap}

We represent the octahedron with centroid $C_1=(0,2,0,0)$ in $\R^4$ by the $4\times 6$ matrix
$$f_1=\left[\begin{array}{cccccc}
2 & -2 & 0 & 0 & 0 & 0 \\
 2 & 2 & 2 & 2 & 2 & 2 \\
 0 & 0 & 2 & -2 & 0 & 0 \\
 0 & 0 & 0 & 0 & 2 & -2 \\
\end{array}\right]\,.$$
Notice that the dot product of $C_1$ with every column of $f_1$ is 4, so the columns
of $f_1$ are the vertices of this octahedral facet. The first unfolding (linear) map 
$$M_1  \ = \ \left[\begin{array}{cccc}
-6 & 0 & -48 & -24 \\
 -48 & 0 & -6 & 24 \\
 24 & 216 & -24 & 42 \\
\end{array}\right]$$
transforms the first octahedral facet $f_1$ in $\R^4$ into a regular octahedron $M_1 f_1$ in $\R^3$
$$M_1 f_1  \ = \ 4\left[\begin{array}{cccccc}
-3 & 3 & -24 & 24 & -12 & 12 \\
 -24 & 24 & -3 & 3 & 12 & -12 \\
 120 & 96 & 96 & 120 & 129 & 87 \\
\end{array}\right].$$
One can verify that these columns coincide with the vertices of a regular octahedron. 
If we perform the analogous computation for the second octahedron in our chain, we see
$$M_2 f_2 \ = \ 4\left[\begin{array}{cccccc}
3 & -24 & 12 & -24 & 12 & -15 \\
 24 & 24 & 15 & -3 & -12 & -12 \\
 96 & 69 & 60 & 96 & 87 & 60 \\
\end{array}\right]\,.$$
Inspecting the columns of $M_1f_1$ and $M_2f_2$, notice they both share the triangle
with vertices $4(3,24,96)$, $4(-24,-3,96)$, and $4(12,-12,87)$. We can continue unfolding in this manner.

We show that a vertex of the 14th octahedron lies interior to the first octahedron in our chain. 
In $\R^3$, this final octahedron is represented by
$$
M_{14} f_{14} \ = \ 4\left[\begin{array}{cccccc}
41 & 16 & 32 & 4 & 20 & -5 \\
20 & -8 & 11 & 25 & 44 & 16 \\
140 & 133 & 104 & 148 & 119 & 112 \\
\end{array}\right]\,.$$
The last column of this matrix may be written as
$$4\left[-15, 6, 112\right]^\top
\ = \ M_1 f_1\cdot \frac{1}{243} \left[1, 108, 14, 6, 113, 1\right]^\top\,.$$
Here, the column vector $(1,108,14,6,113,1)$ represents the coefficients in our convex combination, and its scaling $1/243$ is less than 1, demonstrating overlap of the unfolding.

%
%
\section{The 120-cell} \label{s:120}
\subsection{Quaternion Algebra}

We recall some properties of the 120-cell: it is a 4D convex polytope having 120 identical dodecahedral facets,  720 regular  pentagonal faces, 1200 edges, and 600 vertices. 
To demonstrate the existence of a non-net unfolding of the 120-cell, we present a chain of nine dodecahedral facets whose initial and terminal facets intersect when unfolded.
As with the 24-cell, facets are identified by their centroids and the chain is described via a path in the one-skeleton of the dual 600-cell. 
The vertices of the 600-cell can be realized as the 120 points:
	\begin{itemize}
	\item 8 from permutations of $(\pm 1, \ 0, \ 0, \ 0)$
	\item 16 of the form $(\pm 1/2, \ \pm 1/2, \ \pm 1/2, \ \pm 1/2)$
	\item  96 from even permutations of $(0, \ \pm \psi/2, \ \pm 1/2, \ \pm \phi/2)$.
	\end{itemize}
Here, $\phi= (1+\sqrt{5})/2$ is the golden ratio and $\psi=1/\phi = (-1+\sqrt{5})/2$.

If each vertex $(w,x,y,z)$ is identified with a corresponding unit quaternion $w+xi+yj+zk$, they form a group under quaternion multiplication, the \emph{icosian group} 
(or the \emph{binary icosahedral group}).  For a vertex $v$, the twelve adjacent vertices are given by $v \cdot w$ where $w$ is one of: 
\[ (\phi/2, \ \pm 1/2, \ \pm \psi/2, \ 0), \ (\phi/2, \ 0, \ \pm 1/2, \ \pm \psi/2), \ (\phi/2, \ \pm \psi/2, \ 0, \ \pm 1/2). \]
	
\noindent Hence, we can describe a path that starts at $(1,0,0,0)$ and moves along the edges of the 600-cell 
via a sequence of multiplications. 
One such sequence (the one that is of interest to us) is given in Table~\ref{t:120-list}. 
The fact that there are no repeated coordinates guarantees that this is properly a path (not a loop) on the dual. 

\begin{table}[h]
\begin{tabular}{lrrrr}
Vertex (600-cell) & $w$ & $x$ & $y$ & $z$ \\
\hline \hline
$f_1$ & 1 & 0 & 0 & 0 \\	
$f_2$ & $\phi/2$ & 0 & 1/2 & $\psi/2$ \\
$f_3$ & 1/2 & 1/2 & 1/2 & 1/2 \\
$f_4$ & 0 & $\psi/2$ & 1/2 & $\phi/2$ \\
$f_5$ & $-1/2$ & 0 & $\psi/2$ & $\phi/2$ \\
$f_6$ & 0 & $-\psi/2$ & 1/2 & $\phi/2$ \\
$f_7$ & 1/2 & 0 & $\phi/2$ & $\psi/2$ \\
$f_8$ & $\phi/2$ & $\psi_2$ & 0 & $1/2$ \\
$f_9$ & 1/2 & $\phi/2$ & 0 & $\psi/2$ 
\end{tabular}\vspace{.2in}
\caption{A sequence of nine vertices along a path in the 600-cell.}
\label{t:120-list}
\vspace{-.2in}
\end{table}

\subsection{Dodecahedral Geometry}

The twenty closest vertices to $(1,0,0,0)$ are the vertices of the first dodecahedral facet. After scaling by a factor of $16/\phi$ to give coordinates in the ring $\Z[\sqrt{5}]$, they are 
	\begin{itemize}
	\item 8 of the form $(2+4\phi,\ \pm 2,\ \pm 2, \ \pm 2)$
	\item 4 of the form $(2+4\phi,\  \pm 2\psi, \ \pm 2\phi, \ 0)$
	\item 4 of the form $(2+4\phi,\ \pm 2\phi, \ 0, \ \pm 2\psi)$
	\item 4 of the form $(2+4\phi,\ 0, \ \pm 2\psi,\  \pm 2\phi)$.
	\end{itemize}
Once vertices of this first facet are calculated, those of the subsequent eight facets of the chain can be 
found by quaternion multiplication.  

We unfold these into three-dimensional space as follows: The first facet $f_1$ evidently lies 
in the three-dimensional subspace corresponding to a real part of $2 + 4\phi=4+2\sqrt{5}$, so it can be projected by 
truncating the real coordinate, represented by the matrix	
	\[ M_1 = \begin{bmatrix} 0 & 1 & 0 & 0 \\ 0 & 0 & 1 & 0 \\ 0 & 0 & 0 & 1 \end{bmatrix}. \]	

\begin{figure}[h]
\includegraphics[width=.5\textwidth]{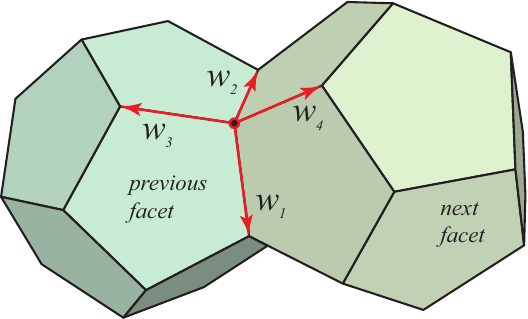}
\caption{Moving from one dodecahedral facet to another.}
\label{f:120-attach}
\end{figure}

To calculate the second projection matrix, choose one of the five vertices on the face shared by facets $f_1$ and $f_2$. Four edge vectors emanate from it: two ($v_1$ and $v_2$) are on the shared face, a third ($v_3$) is an edge of $f_1$, and the fourth ($v_4$) is an edge of $f_2$. The projections $w_1$, $w_2$, $w_3$ of $v_1$, $v_2$, $v_3$ are known from the previous projection. The projection of $v_4$ is the reflection of $w_3$ across the plane containing $w_1$ and $w_2$, which we call $w_4$; see Figure~\ref{f:120-attach}.

\renewcommand\arraystretch{1.6}

The vectors $v_1$, $v_2$, and $v_4$ span a three-dimensional subspace of $\R^4$ containing the second facet. Extend to a basis for $\R^4$ with one additional basis vector $v_0$, orthogonal to the previous three. The projection of the second facet is given by the linear transformation
	\[ T: \R^4 \to \R^3: v_1 \mapsto w_1, \ v_2 \mapsto w_2, \ v_4 \mapsto w_4, \ v_0 \mapsto 0. \]
The second projection matrix is
$$M_2 = 
\begin{bmatrix} 0 & 1 & 0 & 0 \\ 
-\frac{1}{2} & 0 & \frac{3}{4} + \frac{\sqrt{5}}{20}  & \frac{1}{4} - \frac{3\sqrt{5}}{20}  \\
\frac{1}{4} -\frac{\sqrt{5}}{4} & 0 & \frac{1}{4} -\frac{3\sqrt{5}}{20} & \frac{1}{2} + \frac{\sqrt{5}}{5}.
\end{bmatrix}
$$

\renewcommand\arraystretch{1.3}

\subsection{The Overlap}

\begin{figure}[h]
\includegraphics[width=\textwidth]{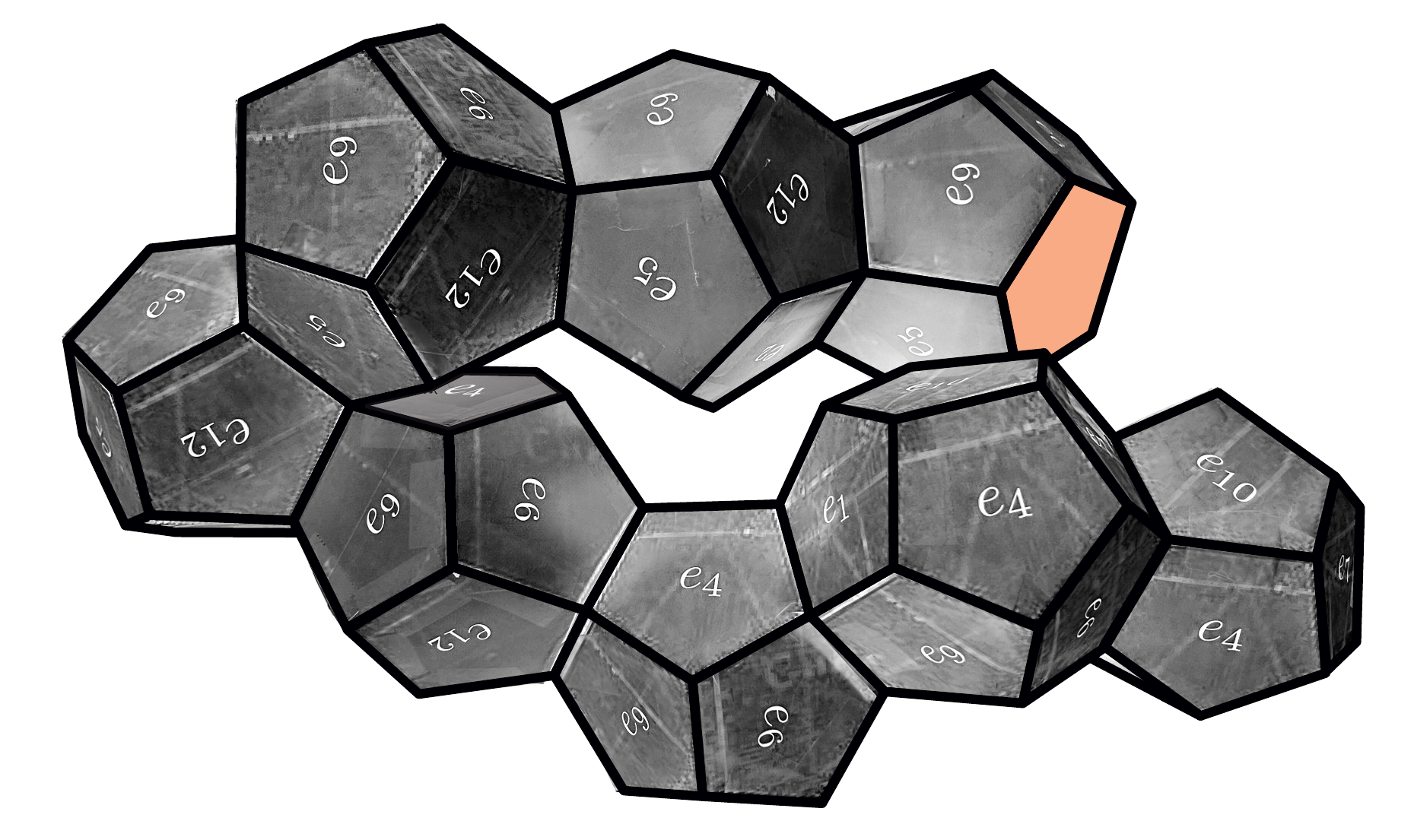}
\caption{Chain of eight dodecahedra in the partial unfolding of the 120-cell.  The addition of a dodecahedron along the marked pentagonal-face results in an overlap.}
\label{f:120-paper}
\end{figure}

We proceed in this way, unfolding a chain of nine facets in the 120-cell.  Figure~\ref{f:120-paper} shows a photograph of eight dodecahedra made of paper, held together by tape. 
Thick black markings have been drawn on top of this photograph to clearly delineate the boundary edges.  Here, when the final facet $f_9$ is attached along the shaded pentagonal face, it will intersect the first facet $f_1$ in this chain. To be precise, one of the projected vertices of $f_9$,
$$v = \left ( \frac{14}{5}, -\frac{6}{5} + \frac{8\sqrt{5}}{25}, -\frac{2}{5} + \frac{16\sqrt{5}}{25} \right ),$$
lies within the tetrahedron whose four corners are the vertices of $f_1$ that have the forms 
$$\{(2, 2, 2), \ (2, -2, 2),\  (2\phi, 0, 2\psi),\  (2\phi, 0, -2\psi)\}\,,$$
and hence lies in the larger dodecahedron that contains it. This can be confirmed by computing equations of the planes of its four faces and comparing the value of $v$ in those functions to that of the origin.

%
%
\section{The 600-cell} \label{s:600}
\subsection{Matrix Coordinates}

\begin{figure}[h]
\includegraphics[width=\textwidth]{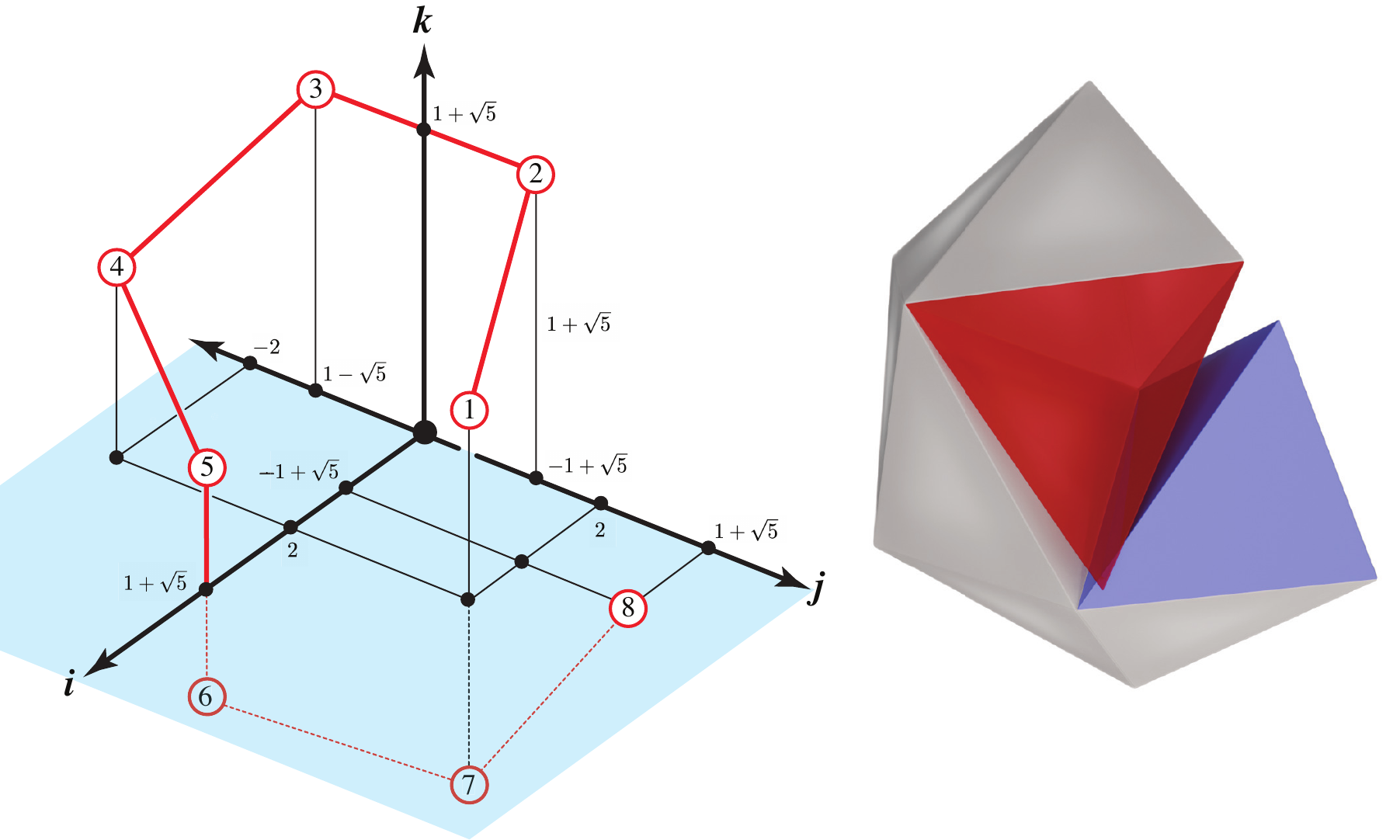}
\caption{The left side shows the path given in Table~\ref{t:600}, whereas the right is the unfolding of the eight tetrahedra of a chain in the 600-cell (starting at the red cell and moving to the purple).}
\label{f:600}
\end{figure}

We recall some properties of the 600-cell: it is a 4D convex polytope having 600 identical tetrahedral facets,  1200 equilateral triangular faces, 720 edges, and 120 vertices. 
Although the all-net failure of the 600-cell was established \cite[Theorem 11]{dh}, we include it here both for completeness, and newly reframed in the unfolding language above. The 600-cell facets in our chain share
a common vertex, corresponding to a path along the edges of a single dodecahedral facet in the dual 120-cell. 
As before, we first calculate the vertex 
coordinates of the path along the dodecahedral facet as shown in \cite[Figure 7]{dh}, given in 
Table~\ref{t:600} below. There are no repeated coordinates, so this 
describes a path (and not a loop) on the dual. The left side of Figure~\ref{f:600} shows 
the path and the eight labeled vertices.

\begin{table}[h]
\begin{tabular}{lrrrr}
Vertex & $w$ & $x$ & $y$ & $z$ \\
\hline \hline
$1$ & $4+2\sqrt{5}$ & $2$ & $2$ & $2$ \\	
$2$ & $4+2\sqrt{5}$ & $0$ & $-1+\sqrt{5}$ & $1+\sqrt{5}$ \\
$3$ & $4+2\sqrt{5}$ & $0$ & $1-\sqrt{5}$ & $1+\sqrt{5}$ \\ 
$4$ & $4+2\sqrt{5}$ & $2$ & $-2$ & $2$ \\
$5$ & $4+2\sqrt{5}$ & $1+\sqrt{5}$ & 0 & $-1+\sqrt{5}$ \\
$6$ & $4+2\sqrt{5}$ & $1+\sqrt{5}$ & 0 & $1-\sqrt{5}$ \\
$7$ & $4+2\sqrt{5}$ & $2$ & $2$ & $-2$ \\
$8$ & $4+2\sqrt{5}$ & $-1+\sqrt{5}$ & $1+\sqrt{5}$ & 0 
\end{tabular}
\bigskip
\caption{Vertex coordinates of a path along a dodecahedral facet.}
\label{t:600}
\vspace{-.2in}
\end{table}

Each is a centroid $C$ of a tetrahedral facet whose vertices $v$ are the minimum distance away, given by $|C - v|^2 = 336-144\sqrt{5}$. For instance, the tetrahedron centered at the first point has four vertices 
in Table~\ref{t:4vert}.

\begin{table}[h]
\begin{tabular}{lrrrr}
Vertex & $w$ & $x$ & $y$ & $z$ \\
\hline \hline
$v_0$ & $-8+8\sqrt{5}$ & 0 & 0 & 0 \\
$v_1$ & $8$ & 0 & $-4+4\sqrt{5}$ & $12-4\sqrt{5}$ \\
$v_2$ & $8$ & $12-4\sqrt{5}$ & 0 & $-4+4\sqrt{5}$ \\
$v_3$ & $8$ & $-4+4\sqrt{5}$ & $12-4\sqrt{5}$ & 0 
\end{tabular}	
\bigskip
\caption{Four vertices of the tetrahedron centered at the first vertex of Table~\ref{t:600}.}
\label{t:4vert}
\vspace{-.2in}
\end{table}

\noindent Note that subsequent facets differ from the previous by only one vertex, so coordinates of the new vertex are easily determined from the three other vertices and the centroid. 

\subsection{The Overlap}

Finally, we unfold this tetrahedral chain into the 3D plane $x+y+z+w=1$ in $\R^4$, mapping the first tetrahedron in the chain as follows:
	\[ v_0 \mapsto w_0 = (1,0,0,0), \quad v_1 \mapsto w_1 = (0,1,0,0), \]
	\[  v_2 \mapsto w_2 = (0,0,1,0), \quad v_3 \mapsto w_3 = (0,0,0,1). \]

\begin{rem}
This is not an isometry: the length of an edge of the image is $\sqrt{2}$, which is not the length of an edge of the original. However, it scales all distances equally, so the image of a regular tetrahedron will still be a regular tetrahedron.
\end{rem}

The second facet shares three of four vertices with the first. The new vertex can be found by reflecting $w_3$ across the shared face $F$:  
	\[ w_3 + 2(C-w_3) = (2/3,2/3,2/3,-1)  \]
 where $C=(w_0+w_1+w_2)/3$ is the centroid of $F$. Continuing in this way, the vertices of the final projected facet are:

\bigskip
\begin{center}
\begin{tabular}{ll}
$w_0 \ = \ (1,0,0,0)$ & \ \ \ \ $w_1 = \frac{1}{3^4}(4,-11,4,84)$ \\ \\
$w_2 = \frac{1}{3^7} (20, 2132, -385,420)$ & \ \ \ \ $w_3 = \frac{1}{3^6} (550,310, -746, 615)$.\\
\end{tabular}
\end{center}
\bigskip

\noindent Now observe that the first facet in this chain corresponds exactly to the portion of 
$x+y+z+w=1$ where all four coordinates are positive. The point 
\begin{equation*}
(17/20)w_1 +  (3/20)w_2 
= \frac{1}{3^6 \cdot 20} (632,449,227,13272),
\end{equation*}
which lies along the $w_1w_2$ edge of the last facet, has all positive coordinates and hence lies in the first facet.	
This overlap is displayed on the right side of Figure~\ref{f:600}.

%
%
\bibliographystyle{amsplain}

\end{document}